\documentclass{article}

\usepackage{arxiv}

\usepackage[utf8]{inputenc}
\usepackage[T1]{fontenc}
\usepackage{hyperref}
\hypersetup{pdfborder={0 0 0}}
\usepackage{url}
\usepackage{tabularx}
\usepackage{longtable}
\usepackage{booktabs}
\usepackage{amsfonts}
\usepackage{nicefrac}
\usepackage{microtype}
\usepackage{multirow}
\usepackage{cleveref}
\usepackage{lipsum}
\usepackage{graphicx}
\usepackage{float}
\usepackage{array}
\usepackage[numbers]{natbib}
\usepackage{doi}
\newcolumntype{P}[1]{>{\raggedright\arraybackslash}p{#1}}
\hypersetup{
pdftitle={Method for aggregating unstructured data using LLM},
pdfsubject={LLM, Data Aggregation, Unstructured Data Processing, Embedding, Auto Validation, JSON},
pdfauthor={Lazebnyi Vsevolod, Tereshkina Natalia, Maria Shabarina, Dmitriy Fedorov},
pdfkeywords={LLM, Data Aggregation, Unstructured Data Processing, Embedding, Auto Validation, JSON},
}

\title{Method for aggregating unstructured data using LLM}

\author{ 
	\href{https://orcid.org/0009-0002-2954-7280}{\includegraphics[scale=0.06]{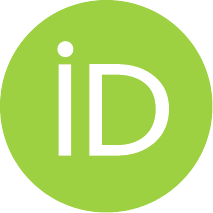}\hspace{1mm}Lazebnyi Vsevolod}, \href{https://orcid.org/0009-0009-5674-9989}{\includegraphics[scale=0.06]{orcid.pdf}\hspace{1mm}Tereshkina Natalia}, \href{https://orcid.org/0009-0004-8737-8520}{\includegraphics[scale=0.06]{orcid.pdf}\hspace{1mm}Maria Shabarina}, \href{https://orcid.org/0000-0003-3462-8392}{\includegraphics[scale=0.06]{orcid.pdf}\hspace{1mm}Dmitriy Fedorov} \\
	ITMO University, Saint-Petersburg, Russia\\
	\texttt{vllazebnyi@gmail.com; tereshkina2004@mail.ru; maria12262004@yandex.ru; dafedorov@itmo.ru} \\
}

\date{}

\begin{document}
\maketitle

\begin{abstract}
This paper presents a method for the automated collection and aggregation of unstructured data from diverse web sources, utilizing Large Language Models (LLMs). The primary challenge with existing techniques is their instability when the structure of webpages changes, their limited support for dynamically loaded content during information collection, and the requirement for labor-intensive manual design of data pre-processing processes. The proposed algorithm integrates hybrid web scraping (Goose3 for static pages and Selenium+WebDriver for dynamic ones), data storage in a non-relational MongoDB database management system (DBMS), and intelligent extraction and normalization of information using LLMs into a predetermined JSON schema. A key scientific contribution of this study is a two-stage verification process for the generated data, designed to eliminate potential hallucinations byy comparing the embeddings of multiple LLM outputs obtained with different temperature parameter values, combined with formalized rules for monitoring data consistency and integrity. The experimental findings indicate a high level of accuracy in the completion of key fields, as well as the robustness of the proposed methodology to changes in web page structures. This makes it suitable for use in tasks such as news content aggregation, monitoring, and log analysis in near real-time mode, with the capacity to scale rapidly in terms of the number of sources.
\end{abstract}


\section{Introduction}
To solve the problem of aggregating unstructured data from streaming sources, we propose a technique that automates the process of extracting and organizing web data using large language models (LLMs) \citep{ref1}.

Extracting information from diverse and dynamically changing web sources, including news portal sites, online marketplaces, and system logs, is becoming an essential task for business intelligence, market monitoring, and competitive analysis. Traditional data extraction methods based on XPath/CSS selectors or specialized parsing tools are used for various purposes, such as monitoring prices from competitors, reviewing links on a webpage, transferring product details, and so on. However, they continue to be used to collect data from sources that do not have a permanentlyly predictable structure, such as media news articles and event announcements.

In its raw form, this data does not have a common structure and requires more detailed processing. Standard collection and aggregation algorithms are often inadequate in such scenarios, as they are sensitive to changes in source data, necessitating additional time and effort to support parsing of each source and perform post-processing operations.

The development of large language models (LLM) has provided novel opportunities for automating data extraction \citep{ref2}. These models can interpret the context of a given text and generate structured responses, making them suitable for various tasks \citep{ref3}.

The integration of LLMs with retrieval-augmented generation (RAG) techniques and hybrid pipelines, such as ScrapeGraphAI and LangChain, enhances the flexibility, adaptability, and scalability of these systems \citep{ref4}. This integration also reduces the need for manual configuration when dealing with static and dynamic web pages.

However, applying LLMs directly to each page may result in high computational costs, and it does not solve the issues of data duplication and validation.

Recent research in this area has led to the development of hybrid algorithms that integrate classical parsing techniques, deduplication using semantic embeddings \citep{ref5}, intelligent data structuring with large language models (LLM), and two-stage verification of the generated data to prevent hallucinations – incorrect data output formats \citep{ref6}. This approach enhances the system's resilience to changes in the source data, ensuring high-quality and scalable solutions for processing unstructured web content in real-time. The effectiveness of this approach has been validated through experimental results in various subject areas.

The shift towards LLM-based and hybrid architectures represents a promising direction for building sustainable and scalable systems capable of automatically extracting and organizing web data.

\section{Overview of existing approaches}
\subsection{Comparison of approaches}
The various approaches and their methods are similar in intent, but at the same time, they have their own advantages and disadvantages. Table~\ref{tab:comparison} below outlines a selection of methods that can be employed when carrying out similar tasks of collating unstructured data during streaming processing.

\begin{table}[htbp]
\caption{Comparison of specified tools and approaches.}
\centering
\renewcommand{\arraystretch}{1.15}

\begin{tabularx}{\textwidth}{|p{2.6cm}|X|X|X|}
\hline
\textbf{Approaches} & \textbf{Method} & \textbf{Advantages} & \textbf{Disadvantages} \\
\hline

\multirow{5}{=}{\parbox{2.4cm}{Parsing}}
& XPath/CSS-selectors & High speed, simplicity, low cost & Fragility to structure changes, does not work with dynamic content \\
\cline{2-4}
& Regular expressions & Flexibility, precision on known patterns & Requires manual configuration, does not transfer well \\
\cline{2-4}
& Headless browsers (Selenium, Playwright) & Work with dynamic pages & High cost, low speed, complexity of support \\
\cline{2-4}
& Template methods & Extraction automation, suitable for similar pages & Limited versatility, sensitivity to structure \\
\cline{2-4}
& Manual template generation & High accuracy on small volumes & Labor-intensive, does not scale \\
\hline

\multirow{4}{=}{\parbox{2.4cm}{Information extraction (IE)}}
& CRF, BERT, transformers & High accuracy, resistance to local changes & Requires markup, low portability \\
\cline{2-4}
& Ontological methods & Deep understanding, extensibility & Requires expert, complex support \\
\cline{2-4}
& Lexical patterns & Easy implementation, transparency & Limited flexibility \\
\cline{2-4}
& Unsupervised methods & No markup needed, automation & Less accurate, harder interpretation \\
\hline

\multirow{2}{=}{\parbox{2.4cm}{Structuring using LLM}}
& GPT-4, Llama & Versatility, deep understanding of context & Cost, latency, hallucinations \\
\cline{2-4}
& Integration with ontologies & Improved accuracy & Complex integration \\
\hline

\multirow{1}{=}{\parbox{2.4cm}{Combined approach}}
& Integration of parsing, IE, LLM & Efficiency, accuracy, automation & Complex implementation \\
\hline

\multirow{1}{=}{\parbox{2.4cm}{Data Mining}}
& ML methods & Pattern detection, scalability & Requires large datasets \\
\hline

\end{tabularx}
\label{tab:comparison}
\end{table}

\subsection{Parsing}
Traditional web scraping tools, such as Beautiful Soup and Scrapy, are widely used for extracting data from static HTML pages due to their speed, low cost, and ease of implementation \citep{ref7}. These tools rely on DOM tree analysis, using XPath/CSS selectors, as well as regular expressions, to ensure deterministic results with an unchanged page structure. However, these methods are not practical for websites with dynamically loaded content, such as single-page applications (SPAs) and sites with endless scrolling and JavaScript interactivity. For these types of websites, additional tools like Selenium and Playwright must be used \citep{ref8}. Additionally, any changes to the layout or structure of target blocks can destabilize parsers, requiring manual intervention. This reduces the effectiveness of traditional scraping tools for long-term monitoring of diverse resources.

\subsection{Information Extraction}
Advanced approaches utilize statistical and neural network models for the extraction of entities and relationships \citep{ref9}. These techniques include BiLSTM-CRF, CRF taggers, and modern transformers that are fine-tuned on specialized datasets \citep{ref10}. The main advantage of these approaches is their high accuracy in extracting information from familiar data, as well as their resistance to local text changes. However, such models require marked-up corpora for each new entity type or source. Additionally, the transfer of these models to new categories of data often results in a decrease in quality, limiting their versatility when dealing with highly variable source structures.

\subsection{Structuring using LLM}
With the advent of large language models such as GPT-4, Claude and Llama, it has become possible to extract structured information from unstructured text using prompts, including a description of the target structure, examples and additional transformation rules \citep{ref11}. This approach offers several advantages, including universality, zero- or few-shot learning capabilities, and a deep understanding of contextual meaning in text \citep{ref12}. However, LLM-based methods have some drawbacks, including high resource and latency costs, stochastic outputs (which can lead to ``hallucinations'') and a lack of built-in mechanisms for data deduplication or formal validation of results \citep{ref6}.

\subsection{Approaches combination}
The proposed method combines the strengths of several existing methods: the effectiveness of traditional data collection techniques, the structured representation of entities in information extraction (IE) models, and the flexibility of large language model (LLM) engineering \citep{ref13}.

The limitations of each approach are addressed through pre- and post-processing steps, as well as a two-stage verification process to detect hallucinations and other artifacts, allowing for automated data extraction from complex and evolving sources.

\section{Suggested method}

\subsection{Method}
The suggested method is implemented in the form of a stable and scalable payline consisting of four consecutive stages. It corresponds to modern trends in automating the processing of unstructured data using hybrid approaches, which we will discuss further step by step.

Step 1. Hybrid data collection and initial storage. A combination of tools is used to overcome the limitations of traditional parsing techniques. This includes the use of tools such as BeautifulSoup and Goose3 for the rapid parsing of static web pages, as well as the automation of browser interactions through Selenium and WebDriver for processing dynamic content \citep{ref1}. This approach enables the collection of data from a variety of websites, including single-page applications and pages that dynamically load content. Goose3 utilizes automated selection of CSS selectors to extract data from the primary page, allowing to catch data from the related links and automatically extract desired text. This ensures that a potential change in web page structure does not create the need to change the parsing algorithm.

Step 2. Semantic Deduplication. To eliminate duplicate records and those that are similar in meaning, embeddings are generated (for example, using the Sentence-BERT or the all-MiniLM-L6-v2 models) and calculating the cosine similarity between the new and existing documents \citep{ref5}. If the similarity exceeds a specified threshold, the record is considered a duplicate and discarded, which saves resources and prevents the reprocessing of identical data.

\begin{figure}[htbp]
	\centering
	\includegraphics[width=0.8\textwidth]{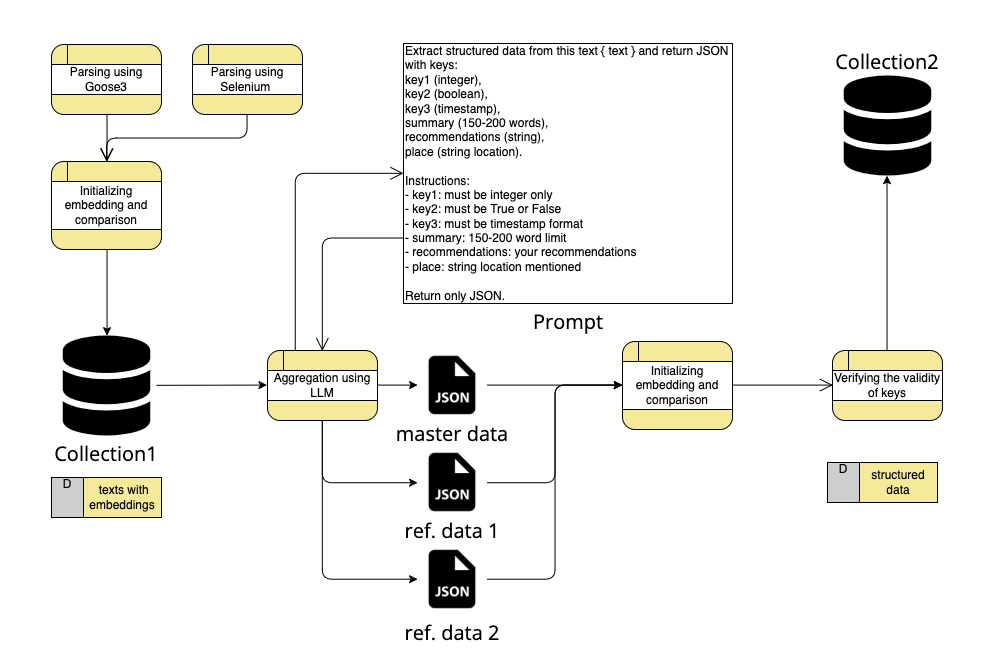} 
	\caption{DFD schema of the method.}
	\label{fig:dfd}
\end{figure}

Step 3. Data extraction and normalization using an LLM. The data of the object is transferred to an LLM API, along with a prompt that contains instructions, target JSON schema details, an example, and processing rules \citep{ref11}. The LLM then extracts key entities from the text, inserting them into the required JSON fields according to a specified template, aggregates and normalizes the data, and returns structured objects, which are stored in a separate collection for ``normalized storage''. The approach described in Figure~\ref{fig:dfd} can be easily adapted for various scenarios, such as generating digests, extracting product features, making recommendations, and more.

Step 4. Multi-level validation and quality assurance. To enhance reliability, a dual-stage hallucination check is used: at the first stage, a triple query to LLM with various temperatures to generate three similar texts, followed by a semantic contrast of embeddings to detect hallucinations of essential fields, as well as a formal rule check (precision of formats, mandatory fields, compliance with standards) \citep{ref6}. At the second stage, automatic verification of data types and adherence to a given format. Documents with inconsistencies or errors are submitted for manual inspection or re-processing. This strategy significantly reduces the likelihood of errors and enhances the quality of the final output, eliminating the need for manual validation of each generated outcome \citep{ref14}.

\section{Architecture and processes}

\subsection{Architecture of method}
The developed method is a multi-level hybrid pipeline that combines classical parsing techniques, modern approaches to semantic deduplication, and the use of LLMs for processing and analyzing unstructured data \citep{ref5}. The proposed method provides high robustness to structural changes in data sources. It operates effectively on both static and dynamically evolving websites. When an exact key cannot be explicitly identified in the text, the LLM automatically infers the appropriate value based on the provided context and the detected subtext. If needed, the method supports horizontal scaling at any stage of the pipeline. This combined approach enables high accuracy and adaptability when working with heterogeneous information \citep{ref12}.

\subsection{Method processes}
\begin{figure}[htbp]
	\centering
	\includegraphics[width=0.8\textwidth]{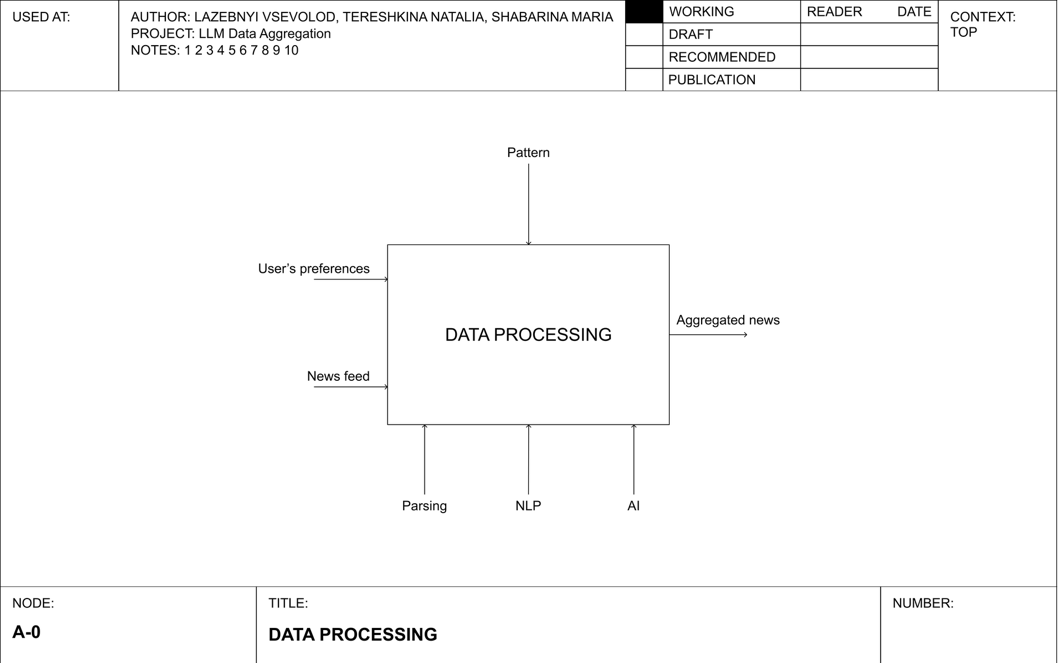} 
	\caption{IDEF0 of method.}
	\label{fig:idef0}
\end{figure}

Next there is an illustration of the operation for method through representative business processes. As a canonical example, we consider the workflow for processing news articles obtained from media web pages. In Figure~\ref{fig:idef0}, we analyze a scenario in which a user must handle a continuous stream of texts originating from heterogeneous and structurally diverse sources, identify key informational elements or generate an abstractive summary of each article while preserving its essential semantic content.

The pipeline produces a structured, semantically enriched, and aggregated representation of news content, suitable for downstream personalized delivery to end users.

\begin{figure}[htbp]
	\centering
	\includegraphics[width=0.8\textwidth]{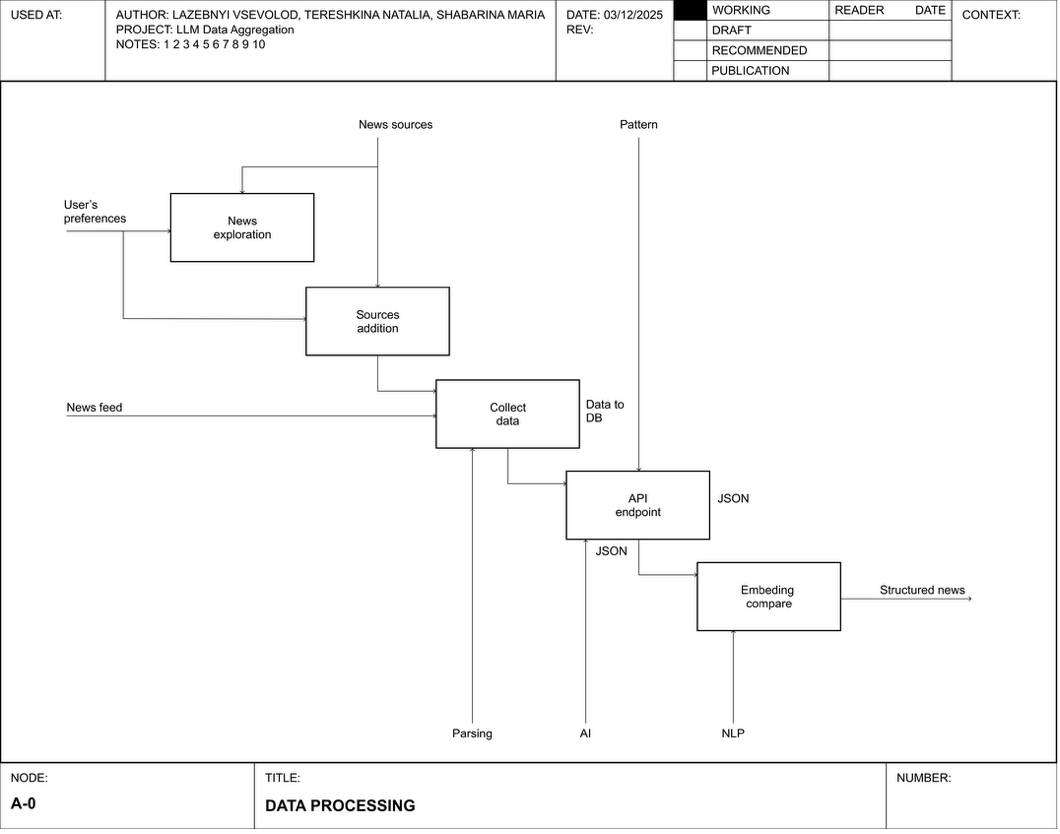} 
	\caption{IDEF1 of method.}
	\label{fig:idef1}
\end{figure}

To enable a deeper examination of the internal data-transformation mechanisms, we further decompose the primary processing workflow in Figure~\ref{fig:idef1}. The IDEF1 diagram provides a detailed specification of the data flows and interactions among the system’s core modules, capturing the end-to-end progression from raw data acquisition to the generation of the final structured response exposed via the API.

The process begins with the ``Collect Data'' stage, where the system acquires news content from predefined or dynamically added sources (Sources Addition). The retrieved unstructured data are then persisted in the database (Data to DB) for subsequent processing. The ``Parsing'' module performs the initial extraction of textual and meta-information, preparing the data for deeper semantic analysis.

The core analytical procedures take place within the NLP and AI modules, which guided by user preferences and predefined processing patterns, conduct fine-grained text analysis. These modules perform entity recognition, topic and sentiment classification, and extraction of key semantic units. A distinctive feature of the proposed approach is the ``Embedding Compare'' stage, where textual data are transformed into vector representations (embeddings) using language models, followed by similarity-based comparison to enable clustering of related news items and content deduplication \citep{ref5}. This transition from a stream of isolated articles to coherent ``Structured News'' significantly enhances downstream usability.

The final step involves producing a structured response through an API endpoint in JSON format, delivering an aggregated and personalized news feed. The entire pipeline supports continuous feedback through dynamic source integration (``Sources Addition'') and is governed by configurable processing patterns (``Pattern''), ensuring system flexibility and adaptability.

Overall, the method exhibits high adaptability, accuracy, and robustness when operating on large volumes of unstructured data, outperforming traditional solutions in both flexibility and output quality.

\section{Experiments and statistics}

\subsection{Experiment 1}
When working with large language models (LLMs), one potential concern is the potential loss of semantic content in the text. To assess the preservation of meaning in text aggregation, the BERT Similarity metric was used, which, unlike superficial metrics such as Levenshtein distance, which assess only symbolic or lexical similarity, allows us to measure the semantic proximity of texts based on their contextual vector representations. This is especially important in aggregation tasks, where the text is restructured and reformulated in JSON format while preserving the meaning for each individual key. So, for the ``summary'' key in JSON, we expect the full meaning of the text to be preserved in an abbreviated format, while for the ``location'' key, only information about the place where the event happened. To verify this, we used the BERT model (bert-base-multilingual-uncased) to compare the collected news articles with the summaries generated from them using key phrases. The average similarity between the two was 87\%, based on a sample of 1600 news articles and generated digests. This indicates that the semantic content was not significantly lost during the process. Figure~\ref{fig:bertsim} below shows an example of data processing based on historical text.

\begin{figure}[htbp]
	\centering
	\includegraphics[width=0.8\textwidth]{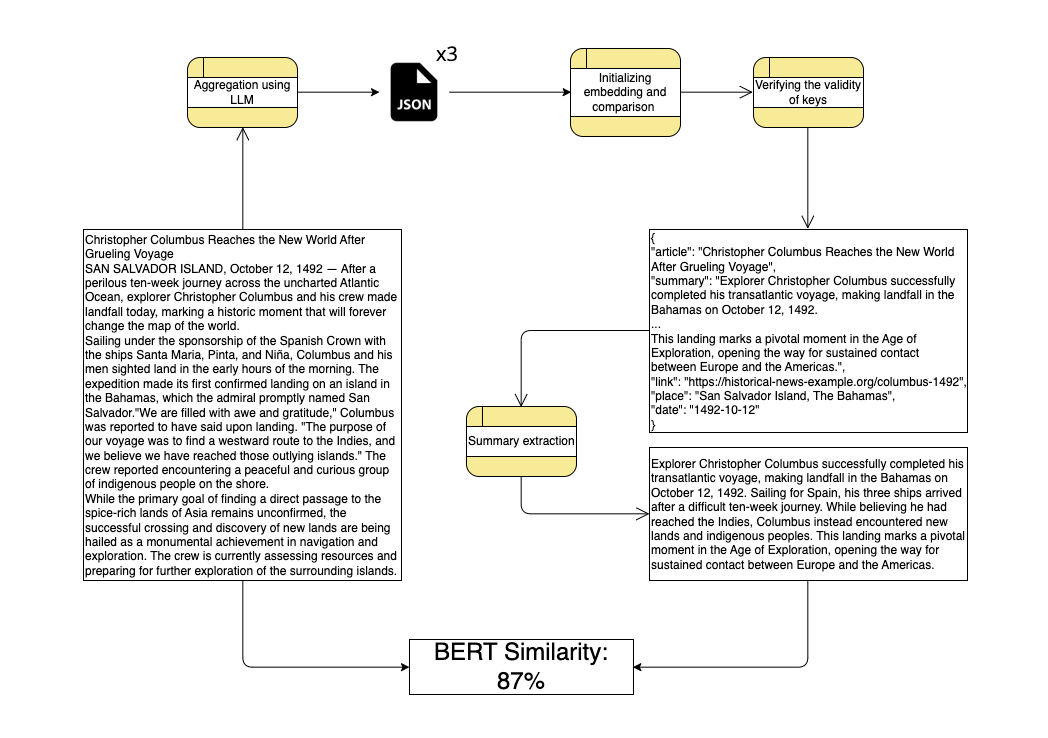} 
	\caption{BERT Similarity between raw text and aggregated text.}
	\label{fig:bertsim}
\end{figure}

The bert-base-multilingual-uncased model was used to calculate BERT Similarity. Vector representations of the texts were obtained by averaging (mean pooling) the hidden states of the last layer of the model over all tokens, except for special service tokens (CLS, SEP), as they perform technical functions and do not carry the semantic load typical of content words and phrases in the text. The similarity between the vectors of the original text and its summary was calculated as the cosine similarity between their normalized vectors. The texts were preprocessed using the standard tokenizer of the BERT model, taking into account its limitations on the maximum length of the input text. For each of these categories the average similarity threshold is between 85\% and 89\%. This number potentially may be increased by enhancing the prompt structure or utilizing the other model for comparison – own model for each category.

\subsection{Experiment 2}
This experiment aims to evaluate the portability of the proposed method to case logs, as well as the effectiveness of the multi-stage strategy in reducing the proportion of hallucinations, while ensuring the correctness of the output JSON structure and acceptable computational costs in terms of both time and token count.

Unlike news articles, log files have a high level of structure and contain technical parameters. They require accurate information extraction, and their structure can vary depending on the source. This makes them unstructured data, but it also presents an opportunity for us to analyze them. A major challenge in processing log files using large language models is the risk of hallucination, such as the generation of non-existent error codes, incorrect IP addresses, and inaccurate timestamps. To manage hallucinations, a multi-stage verification process has been developed based on analyzing the consistency of responses generated using different temperature settings. The process starts with the construction of a baseline response (N0), which has a temperature between 0.3 and 0.7. Additional variants (N1, N2, etc.) with a lower temperature are then generated to reduce the creativity of the model. For each variant, the semantic similarity to N0 is calculated in the embedding space, and the structural consistency of the responses is checked. If the similarity falls below a predetermined threshold, the response is flagged as potentially containing hallucinatory content and either sent for further processing or manual verification.

A sample of 1,200 log files was used in the experiment. This sample consisted of various types of logs, ranging from simple informational records to complex error messages. Each log file was converted into a JSON object, which followed a strictly defined structure. The structure included the following required fields: time stamp, severity level, service name, error code, error message, root cause, and recommended action.

The experiments also utilized the gpt-4o-mini language model. The context size was set to 4,000 tokens, similar to the previous study. A temperature value of 0.7 was applied to generate the reference response, while a temperature of 0.1 was employed for generating test duplicates.

The results of the experiment showed that the proposed approach successfully minimizes the occurrence of hallucinations. Without validation, the percentage of responses with hallucinations was 18.2\%. Utilizing multi-stage verification with N=3 decreased this figure to 3.8\%, while N=5 reduced it to 1.5\%. Additional results, including token expenses and latency times, are provided in Table~\ref{tab:multistage}.

\begin{table}[htbp]
\caption{Multi-stage comparison.}
\centering
\renewcommand{\arraystretch}{1.15}

\begin{tabular}{|p{3cm}|p{3cm}|p{3.2cm}|p{2.8cm}|}
\hline
\textbf{Method} & \textbf{Hallucinations, \%} & \textbf{Tokens per object} & \textbf{Latency, ms} \\
\hline
Base & 18.2 & 800 & 420 \\
\hline
Multi-stage (N=2) & 7.6 & 1600 & 780 \\
\hline
Multi-stage (N=3) & 3.8 & 3200 & 1320 \\
\hline
Multi-stage (N=5) & 1.5 & 6400 & 2160 \\
\hline
\end{tabular}
\label{tab:multistage}
\end{table}

There is a sublinear relationship between the number of duplicates generated and the quality of hallucination control. This relationship can be approximated using a mathematical function of the form: $y = 2\sqrt{x}$, where $x$ represents the number of duplicates, and $y$ represents the improvement in hallucination reduction compared to the base approach. This indicates the principle of diminishing returns, as each additional iteration brings a smaller increase in quality. The obtained relationship supports the practical recommendation of using 2-3 duplicates as a compromise between the quality and cost of hallucination reduction and minimizing processing costs.

\subsection{Experiment 3}
In the framework of an experimental evaluation on a sample of 45 financial documents, the proposed method and the baseline approach LLM-TKIE (Large Language Model–driven Transferable Key Information Extraction) were compared \citep{ref15}.

The LLM-TKIE method was used as the baseline. This is a multi-step pipeline that starts with text detection and recognition (OCR), followed by the processing of unstructured text by the LLM. The key step is a few-shot information extraction, where the LLM generates structured output (JSON) using prompts with a few annotated examples. The pipeline includes a pre-processing step to ensure the completeness of the information and a post-processing step to validate the generated JSON.

The proposed method is a text-centric pipeline with multi-stage validation. The main idea is to minimize dependence on the original layout by collecting full text in already marked objects (pdf, word-docs) and checking the response for hallucinations.

For the experiment, we used a weighted Field-level (F-score): for each target field, we compared the prediction with the reference markup, then aggregated it by document/dataset with field importance weights. The comparison was performed by data type: details and IDs were compared for exact matches, sums were compared for tolerance after normalization, and text was compared for semantic proximity thresholds. The risk of ``fatal'' errors was reflected in the Critical Error Rate, which is the proportion of documents where at least one critical field (sum/details) is incorrect relative to the reference. The Hallucination Rate was calculated as the percentage of documents in which the method filled in a field that was not present in the original document (where the value was not extracted from the source but was invented).

The proposed method, as seen in Table~\ref{tab:comparison2}, demonstrates an improvement in key metrics over the baseline LLM-TKIE.

\begin{table}[htbp]
\caption{Comparison between the baseline and suggested approach}
\centering
\renewcommand{\arraystretch}{1.15}

\begin{tabular}{|p{5cm}|p{3.5cm}|p{3.5cm}|}
\hline
\textbf{Metrics} & \textbf{LLM-TKIE (baseline)} & \textbf{Suggested method} \\
\hline
Field-level (F-score), \% & 86.1 & 98.7 \\
\hline
Critical Error Rate (CER), \% & 12.4 & 2.5 \\
\hline
Hallucination Rate, \% & 28.7 & 3.5 \\
\hline
\end{tabular}

\label{tab:comparison2}
\end{table}

The proposed method with multi-stage validation significantly increases the reliability of data extraction, significantly reducing the level of critical errors and hallucinations while maintaining high accuracy.

\subsection{Statistics}
Table~\ref{tab:existing} presents the data on various methods for the task of value summarization. A similar comparison was conducted in a recent academic paper.

\renewcommand{\arraystretch}{1.15}
\setlength{\tabcolsep}{3pt}
\footnotesize

\begin{longtable}{|P{1.5cm}|P{2cm}|P{2.5cm}|P{2cm}|P{3cm}|P{4cm}|}
\caption{Comparison of existing approaches.}\label{tab:existing} \\
\hline
\textbf{Approach} & \textbf{Processing of duplicates} & \textbf{Similar news detection} & \textbf{Automatisation} & \textbf{Processing loop time} & \textbf{Structure selection / Words limitation} \\
\hline
\endfirsthead
\hline
\textbf{Approach} & \textbf{Processing of duplicates} & \textbf{Similar news detection} & \textbf{Automatisation} & \textbf{Processing loop time} & \textbf{Structure selection / Words limitation} \\
\hline
\endhead
\hline
\endfoot
\hline
\endlastfoot
\citep{ref16} & No & Clusters & Yes & Not stated & LLM + generating three structured summary levels / 4096-token context window, forcing 5--8 sentences per article \\
\hline
\citep{ref6} & No & Specific dataset & Yes & Not stated & Refining LLM / Length is implicitly constrained by the underlying LLM and API usage \\
\hline
\citep{ref17} & No & The study does not cluster or group similar cases. Each case is classified independently & Mostly yes / manual checking & Not stated & LLM prompt / 200k-token context window, allowing full judgments (some over 8{,}000 words) to be processed without truncation \\
\hline
\citep{ref18} & Yes & No & Yes & Not stated & Sentence embeddings and cosine similarity are used. No keyphrase extraction or article structure parsing is performed / Models impose a 512-token input limit for BERT-based systems \\
\hline
\citep{ref19} & Yes & No & Yes & Not stated & The only structure extracted is sentence-level consistency labels (Faithful / Intrinsic / Extrinsic) / Summaries are constrained to about 2--3 sentences \\
\hline
\citep{ref20} & No & No & Yes & Training time is about 500 hours for 350k steps on RTX 2080 Ti & No structural extraction; T5 is trained directly / Inputs and summaries are truncated to 512 tokens \\
\hline
\end{longtable}

Our method demonstrates a novel approach for each of the criteria, which sets it apart from previously developed methods \citep{ref13}.

A key advantage of our system lies in the processing time for the entire news structuring process: from collection to summary. The primary factor influencing this process is the modification of user policy when utilizing LLM, specifically the introduction of restrictions on token count.

\section{Conclusion}
The method is proposed as a subject-independent, experimental and general evaluation in this work focused on news aggregation. The experiments conducted demonstrated a high percentage of retention of the semantic content in the aggregated data, with an average of 87\%. This indicates the preservation of semantic meaning and the absence of significant distortion after processing. The semantic integrity of the text remains intact. Compared to other combined approaches, the proposed method operates in real-time, is resilient to changes in the structure of the source data, and is not affected by variations in the type of data itself. The scientific contribution lies in combining best practices and addressing shortcomings in the aggregation process, without requiring manual verification at each step \citep{ref14}.

The paper proposes a method for integrating and organizing web data that integrates hybrid information gathering, non-relational data storage, language model-based processing \citep{ref1}, and a two-stage verification process to address hallucination issues \citep{ref6}. This technique addresses the key shortcomings of current approaches, including instability in response to layout changes, the requirement for domain-specific training, ongoing system maintenance, and the absence of integrated quality assurance mechanisms.

The experiments and practical implementation of the proposed method in various systems have demonstrated its high versatility and scalability \citep{ref4}. This method can be applied to a wide range of tasks, from the real-time aggregation of news and price data to the seamless analysis of logs and the generation of error correction suggestions. Unlike traditional solutions, this method automatically adapts to new data formats, enabling expedited development of products that rapidly increase the number of information sources.

In future LLM (Large Language Model) development, this algorithm will enhance the quality of data aggregation and will be able to validate the accuracy of processed information.

\end{document}